\pdfoutput=1 

\documentclass{article}
\usepackage[countmax]{subfloat}
\usepackage{ijcai25}

\usepackage{times}
\usepackage{soul}
\usepackage{url}
\usepackage[hidelinks]{hyperref}
\usepackage[utf8]{inputenc}
\usepackage[small]{caption}
\usepackage{graphicx}
\usepackage{amsmath}
\usepackage{amsthm}
\usepackage{booktabs}
\usepackage[switch]{lineno}

\usepackage{float}
\usepackage{overpic}
\usepackage{float}  
\usepackage{subfloat}
\usepackage{stfloats} 
\usepackage[skip=0.5\baselineskip]{caption}
\usepackage{bm}
\usepackage{amsmath}
\usepackage{multirow}
\usepackage{multicol}
\usepackage{enumitem}
\usepackage{subcaption}
\usepackage{threeparttable}
\usepackage{xspace}
\usepackage{color}
\usepackage[normalem]{ulem}
\usepackage{marvosym}   
\usepackage{colortbl}   
\usepackage{tcolorbox}  
\tcbuselibrary{breakable}   
\usepackage{wasysym}    
\usepackage[figuresright]{rotating} 
\usepackage{wrapfig}

\usepackage{xspace}

\newcommand{\eg}{\emph{e.g.,}\xspace}
\newcommand{\ie}{\emph{i.e.,}\xspace}

\title{LLMSeR: Enhancing Sequential Recommendation via LLM-based Data Augmentation}

\author{
    Yuqi Sun$^1$,
    Qidong Liu$^{1,2}$,
    Haiping Zhu$^1$,
    Feng Tian$^1$
    \affiliations
    $^1$Xi’an Jiaotong University,
    $^2$City University of Hong Kong
    \emails
    \{YuqiSun,liuqidong\}@stu.xjtu.edu.cn,
    \{zhuhaiping,fengtian\}@mail.xjtu.edu.cn
}

\begin{document}

\maketitle
\begin{sloppypar}
\begin{abstract}
    Sequential Recommender Systems (SRS) have become a cornerstone of online platforms, leveraging users' historical interaction data to forecast their next potential engagement. Despite their widespread adoption, SRS often grapple with the long-tail user dilemma, resulting in less effective recommendations for individuals with limited interaction records. The advent of Large Language Models (LLMs), with their profound capability to discern semantic relationships among items, has opened new avenues for enhancing SRS through data augmentation. Nonetheless, current methodologies encounter obstacles, including the absence of collaborative signals and the prevalence of hallucination phenomena.
    In this work, we present \textbf{LLMSeR}, an innovative framework that utilizes Large Language Models (LLMs) to generate pseudo-prior items, thereby improving the efficacy of Sequential Recommender Systems (SRS). To alleviate the challenge of insufficient collaborative signals, we introduce the Semantic Interaction Augmentor (SIA), a method that integrates both semantic and collaborative information to comprehensively augment user interaction data. Moreover, to weaken the adverse effects of hallucination in SRS, we develop the Adaptive Reliability Validation (ARV), a validation technique designed to assess the reliability of the generated pseudo items. Complementing these advancements, we also devise a Dual-Channel Training strategy, ensuring seamless integration of data augmentation into the SRS training process.
    Extensive experiments conducted with three widely-used SRS models demonstrate the generalizability and efficacy of LLMSeR.
\end{abstract}

\section{Introduction} \label{sec:introduction}
Recommender Systems (RS) have become ubiquitous across the internet, playing a crucial role in predicting users' potential interests in various items. The aim of the Sequential Recommender Systems (SRS)~\cite{wang2019sequential} is to recommend the next item for users based on their historical interactions. Especially in the context of e-commerce, SRS facilitates online consumption by offering personalized services to individuals~\cite{singer2022sequential}. To capture preferences from users' historical interactions, many SRS models have been proposed and achieved encouraging performance in the past years. For example, recurrent neural networks (RNNs)~\cite{gru4rec2015session}, convolutional neural networks (CNNs)~\cite{tang2018cnn} and attention mechanisms~\cite{lv2019sdm} are applied to extract the users’ long-term interest.

\begin{figure}[t]
\centering
\includegraphics[width=0.9\linewidth]{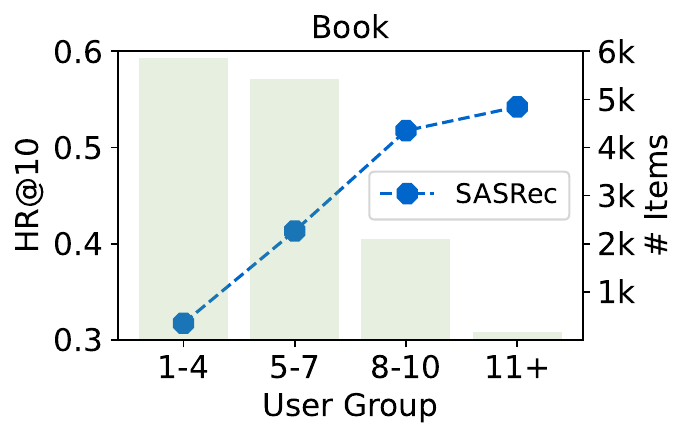}
\caption{The illustration for long-tail user problem.}
\vspace{-3mm}
\label{fig:long_tail}
\end{figure}

Although significant progress has been made in the sequential recommendation, the performance of SRS is weakened by \textbf{Long-tail User Problems}. Specifically, SRS is often good at serving users with abundant historical interactions, while giving worse recommendations to users with fewer historical interactions, \ie the long-tail users. 
As proven in Figure~\ref{fig:long_tail}, above 80\% users only own a few interactions, \ie interaction sequences shorter than 8, but a popular SRS model, SASRec~\cite{kang2018sasrec}, only performs much better for the remaining users who have richer interactions. 
This phenomenon suggests that most users receive unsatisfied recommendation services, which may roughly ruin the user's experience. 
To face such a challenge, data augmentation is a distinct way, whose main idea lies in generating pseudo items for each user. 
At the early stage, \textbf{Co-occurrence-based Methods} prevail, 
which generate pseudo items with the help of the co-occurrence pattern information captured by the general SRS~\cite{transformer2021sequential,ASReP2021augmenting,BiCAT2021sequential}. For example, ASReP~\cite{ASReP2021augmenting} uses the reverse pre-trained SASRec as a booster to iteratively generate pseudo items, treating the top-$1$ results from SASRec as pseudo items. 
However, these approaches only rely on collaborative information, which suffers from noise issues due to inaccurate similarity estimation between users.

Recently, Large Language Models (LLMs) have demonstrated powerful abilities in language comprehension~\cite{openai2023gpt} and complex reasoning~\cite{reason2024language}, driven by training on extensive real-world data. Thus, leveraging the LLMs to alleviate the noise problem from a semantic perspective presents promising. Based on this idea, some works have proposed \textbf{LLM-based methods} for data augmentation~\cite{llama4rec2024integrating,wei2024llmrec}. They use LLMs to understand the user's historical behaviors and item semantics to generate pseudo items. 
However, there are still two major challenges existing in current LLM-based methods. 
(i) \textbf{Missing Collaborative Signals}: LLM-based methods typically rely on LLMs to directly predict pseudo items based on users' historical interactions.
Given the complexity inherent in data augmentation tasks, the accuracy of LLM-based models is constrained by certain limitations. 
Slight variations in language output can lead to the generation of completely different item titles, which may result in irrelevant or out-of-corpus items~\cite{bao2023bi}.
At this point, collaborative information can help address this limitation.
(ii) \textbf{Hallucination}: Most existing works focus on how to use LLMs to generate pseudo items but ignore the defect of hallucination~\cite{hallucination2023survey} of LLMs. It may erode the quality of augmented data. 
These two challenges block the benefits of LLM-empowered data augmentation for the \textbf{Long-tail User Problem} severely.

To address the challenges mentioned above, we propose a \textbf{L}arge \textbf{L}anguage 
\textbf{M}odels Empowered Data Augmentation framework for
\textbf{Se}quential
\textbf{R}ecommendation(\textbf{LLMSeR}).
It is composed of two parts. 
To alleviate the issue of missing collaborative signals, we propose the Semantic Interaction Augmentor (SIA) that combines semantic and collaborative information to jointly enhance user interaction data. Specifically, we first use the pretrained SRS model to generate a candidate pool, and then leverage semantic information to filter out potential noise items from this pool. 
Then, to weaken the impact of hallucination for SRS, we propose the Adaptive Reliability Validation (ARV) module, validating the pseudo items generated by SIA. During training, varying weights are assigned based on the reliability evaluation of each sequence, ensuring that more reliable interactions are prioritized.
In summary, the contributions of this paper are as follows: 

\begin{itemize}[leftmargin=*]
    \item We design a novel LLM-empowered data augmentation framework for sequential recommendation. This framework leverages LLMs to predict user-item interactions from both semantic and collaborative perspectives;

    \item To alleviate the impact of hallucinations, we propose a self-verification and scoring module. To ease the training process, the scoring results will determine the weight between the augmented results and historical interactions;
    
    \item Our method is extensively evaluated on three real-world datasets. The experimental results on three popular SRS models show the generality and effectiveness of LLMSeR.

\end{itemize}
\section{Preliminary} \label{sec:preliminary}

The aim of SRS is to extract user preferences from historical interactions and predict the next item that the user may interact with. We use $\mathcal{U}=\{u_1,u_2,\dots,u_{|\mathcal{U}|}\}$ and $\mathcal{V}=\{v_1, v_2,\dots,v_{|\mathcal{V}|}\}$ to represent the user and item sets.
The historical interaction of user $u$ arranged in timeline can be denoted as $\mathcal{S}_{u}=\{v^{u}_1,v^{u}_2,\dots,v^{u}_{n_u}\}$, where $n_u$ represents the interaction number of $u$. For simplicity, we omit the superscript $u$ in the following sections. The task of SRS is to predict the item of the next interaction given previous $n_u$ interactions. Therefore, the problem can be defined as follows:
\begin{equation}\label{eq1}
   arg \max\limits_{v_i\in \mathcal{V}} P(v_{n_u+1}=v_i|\mathcal{S}_u) 
\end{equation}

\noindent Data augmentation intends to generate pseudo-prior items for users. Specifically, the historical interactions of user $u$ is $\mathcal{S}_u=\{v_1,v_2,\dots,v_{n_u}\}$. The generated pseudo-prior items can be represented as $\mathcal{S}^{pse}_u=\{v_{-M+1},v_{-M+2},\dots,v_{0}\}$, where $M$ represents the number of generated pseudo-prior items. Finally, the augmented sequence of user $u$ can be denoted as $\widetilde{\mathcal{S}}_u=\{v_{-M+1},v_{-M+2},\dots,v_{0},v_1,v_2,\dots,v_{n_u}\}$, which will finally be used for training and inference. 
\section{Method}  \label{sec:method}


\subsection{Framework}
\label{subsec:overview}

\begin{figure*}[t]
\centering
\includegraphics[width=0.9\linewidth]{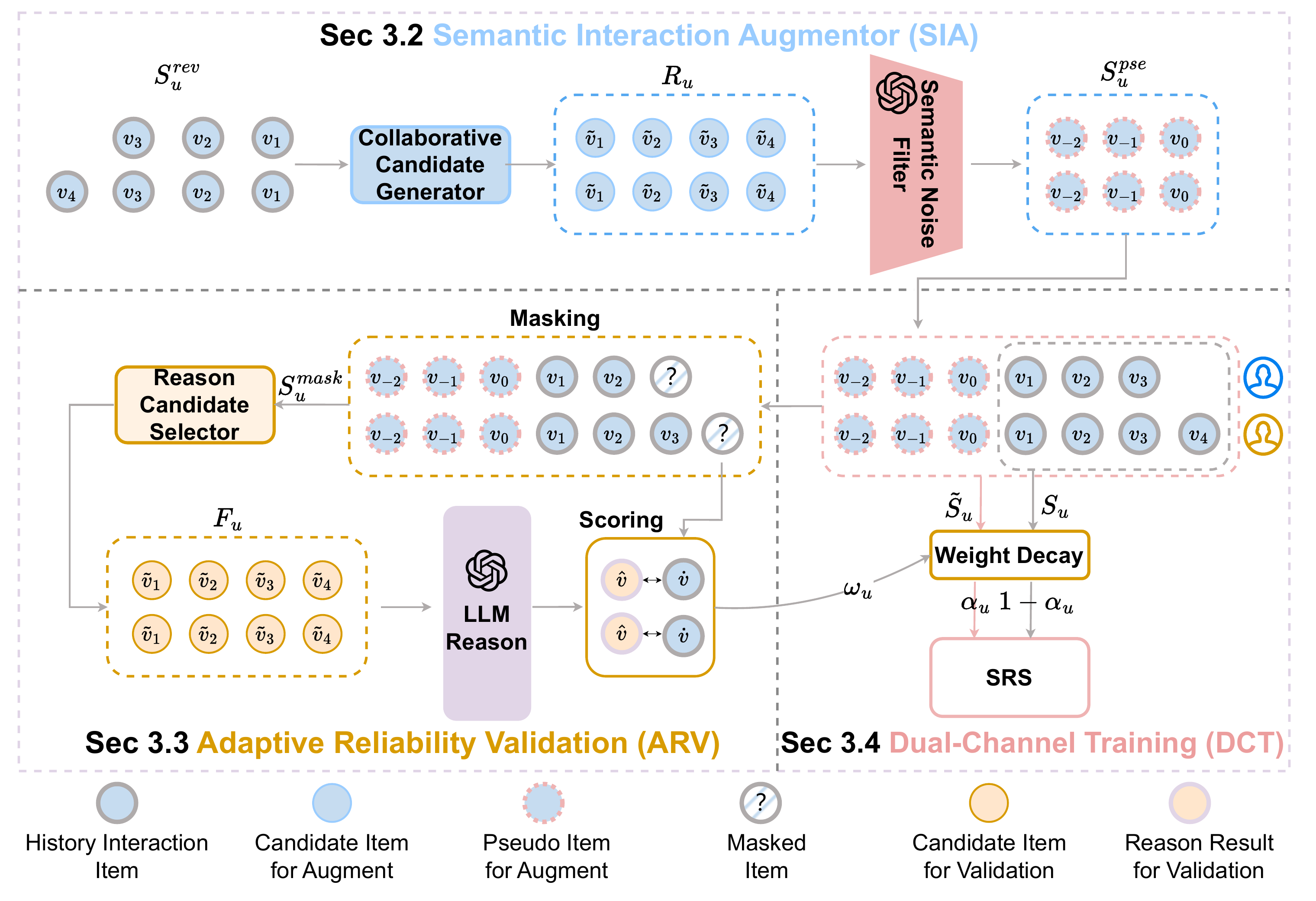}
\caption{The architecture of proposed LLMSeR, illustrated with two users as an example.}
\label{fig:framework}
\vspace{-3mm}
\end{figure*}

As illustrated in Figure~\ref{fig:framework}, we propose an LLMSeR framework for data augmentation in the sequential recommendation. 
It consists of three key components: the Semantic Interaction Augmentor, Adaptive Reliability Validation, and Dual-Channel Training. 
First, the historical interaction sequence of user $u$ is input into the \textbf{Semantic Interaction Augmentor}. This component generates pseudo-prior items based on both semantic and collaborative information, which are then prepended to the beginning of the user's historical interaction sequence. 
Next, the enhanced sequence $\widetilde{\mathcal{S}}_u$ is assigned a corresponding reliability score through the \textbf{Adaptive Reliability Validation} module. This module is designed to assess the relevance of the pseudo-prior items to the user's initial preferences, thereby alleviating the impact of hallucinations from LLMs. 
Finally, we introduce a training strategy called \textbf{Dual-Channel Training}, which adjusts the weights between the historical sequence and the augmented sequence based on their corresponding reliability scores. 
Specifically, if the enhanced sequence $\widetilde{\mathcal{S}}_u$ demonstrates high reliability, it will have a stronger influence on the SRS model during training. Conversely, if the reliability is low, the model will place greater emphasis on the user's original interaction sequence $\mathcal{S}_u$.

\subsection{Semantic Interaction Augmentor (SIA)}\label{subsec:augmentor}

To address the issue of missing collaborative information in LLM-based data augmentation methods, we propose the Semantic Interaction Augmentor. This augmentor consists of a Collaborative Candidate Generator (CCG) and an LLM-based Semantic Noise Filter (SNF).

\noindent \textbf{\textit{Collaborative Candidate Generator (CCG)}}.
Though LLMs can generate interactions from a semantic view, they may face the problem of inefficiency and lack of collaborative signals. Thus, we first use a reverse pre-trained SRS as the CCG to assist the pseudo-prior item generation. The reverse pre-trained SRS can be a general SRS, \eg SASRec~\cite{kang2018sasrec}. In detail, we use CCG to generate the pseudo item candidate pool, so we can incorporate collaborative information embedded in historical interactions into the SIA. Moreover, this effectively filters out items that significantly deviate from the user's original interests. As a result, it enables the LLMs to focus on a smaller, more relevant set of items.
To get the CCG, we organize the user's historical interactions in reverse chronological order. Given the historical interaction sequence $\mathcal{S}_u=\{v_1,v_2,\dots,v_{n_u}\}$, the training data for the SRS is denoted as $\mathcal{S}^{rev}_u=\{v_{n_u},v_{n_u-1},\dots,v_1\}$. 
By such reverse training, the CCG can predict the potential pseudo-prior items for the sequence $\mathcal{S}_u$.
The process of generating the pseudo item candidate pool $\mathcal{R}_u$ based on user historical interactions can be formally described as:
\begin{equation}\label{eq2}
    \mathcal{R}_{u}=f(\mathcal{S}_u^{rev})
\end{equation}
where $\mathcal{R}_{u}=\{\tilde{v}_1,\tilde{v}_2,\dots,\tilde{v}_N\}$ is the set of candidate pseudo items for user $u$. Essentially, it corresponds to the top-$N$ results of the CCG mentioned earlier, where $N$ denotes the size of the candidate pool. The function $f(\cdot)$ represents the process of generating the top-$N$ items using CCG.

\noindent \textbf{\textit{Semantic Noise Filter (SNF)}}. 
To alleviate the noise issue caused by collaborative signals, we leverage LLMs as the semantic noise filter to screen items in the candidate pool. This filtering process is based on the semantic relevance between the user's historical interactions and the items in the candidate set. 
Thus, this approach effectively combines the strengths of both semantic and collaborative information in data augmentation. In detail, we provide the LLMs with auxiliary information (\eg item titles) and the candidate pools derived from the CCG. 
The process of semantic noise filtering can thus be formally described as follows:

\begin{equation}\label{eq3}
    \mathcal{S}_u^{pse}=\text{LLM}(\mathcal{S}^{rev}_u,\mathcal{R}_u)
\end{equation}
Here, $\mathcal{S}_u^{pse}$ represents the set of $M$ pseudo-prior items generated for user $u$. $\text{LLM}(\cdot)$ refers to the filtering process carried out by the large language model, which screens out potential noise from the candidate pool of $N$  items, ultimately selecting $M$ pseudo-prior items for data enhancement.

Specifically, the prompt for generating pseudo items consists of four components:
1) task description, 2) the user's historical interactions, 3) items in the candidate pool, and 4) output format specifications. 
        
        
\noindent For example, \ul{\textless His Item$1$\_title\textgreater} will be replaced by the title of the first interaction item in the user's reverse interaction sequence $\mathcal{S}_u^{rev}$, and \ul{\textless Cand Item$1$\_title\textgreater}  will be replaced by the title of the first item in the pseudo item candidate pool.

Using Semantic Interaction Augmentor, we generate the pseudo-prior items $\mathcal{S}_u^{pse}$. These pseudo items are inserted at the beginning of the user's original historical interaction sequence to form the enhanced interaction sequence $\widetilde{\mathcal{S}}_u$. 

\subsection{Adaptive Reliability Validation (ARV)}\label{subsec:validation}

In this section, we will introduce the proposed Adaptive Reliability Validation, a module designed to alleviate the impact of LLM hallucinations on SRS models. It primarily consists of a Masking module, the Reason Candidate Selector (RCS), an LLM Reason module and a Scoring module.


This approach is based on the assumption that if the pseudo items used for enhancement are consistent with the user's original interests, the interaction enhancement can be deemed reliable. 
Thus, 
the ARV process can generally be divided into three steps: masking, prediction, and evaluation. 
We first use masking to drop a specific historical interaction. Then, in the RCS and Reason components, we attempt to predict the masked item using the enhanced sequence. 
Finally, the scoring module is applied to assess the reliability of each user's enhanced sequence, which is essentially determined by the similarity between the predicted results and the masked items.

\noindent \textbf{\textit{Masking}}. In this process, the most recent historical interaction $v_{n_u}$ of $\mathcal{S}_u$ is masked, so the original $\mathcal{S}_u$ is transformed into $\mathcal{S}_u^{mask}$. 
Specifically, the masked item is denoted as $\dot{v}$ 
serving as the ground truth in the scoring stage. 

\noindent\textbf{\textit{Reason Candidate Selector (RCS)}}. Directly using LLMs for reasoning may overlook the user’s collaborative information. Therefore, we incorporate RCS to assist LLM reasoning. At the same time, this approach can also significantly enhance the reasoning efficiency of LLMs by narrowing the focus to a smaller set of items. Specifically, we employ a forward pre-trained SRS as the RCS to generate the inference candidate pool $\mathcal{F}_u$ based on historical interactions. The term ``forward'' refers to the fact that the RCS is trained using $\mathcal{S}_u$. Formally, this process can be expressed as:

\begin{equation}\label{eq4}
   \mathcal{F}_u=g(\mathcal{S}_u^{mask})
\end{equation}
Here, $g(\cdot)$ denotes the process of generating the inference candidate pool using RCS, and $\mathcal{F}_u=\{\tilde{v}_1,\tilde{v}_2,\dots,\tilde{v}_{H}\}$ refers to the inference candidate pool of user $u$. Essentially, it corresponds to the top-$H$ results of the RCS.

\noindent\textbf{\textit{LLM Reason}}. 
To leverage the abductive reasoning capabilities~\cite{xu2023reasoner} of the LLMs, we guide the reasoning of the LLMs through carefully designed prompts. Specifically, given the enhanced interaction sequence $\mathcal{S}_u^{mask}$ of user $u$, the prompt is designed to ask, like ``What will be the masked item?'' 
The process of guiding LLMs to infer the masked item $\dot{v}$ using $\mathcal{S}_u^{mask}$ can be formally expressed as:

\begin{equation}\label{eq5}
    \hat{v}=\text{LLM}(\mathcal{S}_u^{mask},\mathcal{F}_u)
\end{equation}

\noindent Here, $\text{LLM}(\cdot)$ refers to the process where the LLMs perform reasoning and ultimately infers the most likely masked item $\hat{v}$ from the set of $H$ candidates. In the prompt, we provide the LLMs with the masked enhanced interaction sequence $\mathcal{S}_u^{mask}$ for each user, along with the text description of each item. The detailed prompts used are provided in \textbf{Appendix A.2}.

\noindent\textbf{\textit{Scoring}}. The purpose of the scoring module is to calculate the similarity between the item $\hat{v}$, inferred by the LLMs, and the masked item $\dot{v}$. 
Specifically, we use a pre-trained language model, \ie BERT~\cite{bert2018pre}. 
This process can be formally expressed as:
\begin{equation}\label{eq6}
    \omega_u={\rm cos}(\text{Bert}(\hat{v}),\text{Bert}(\dot{v}))
\end{equation}
Here, $\omega_u$ indicates cosine similarity between the title of the predicted item $\hat{v}$ and the title of the masked item $\dot{v}$.

Finally,
through ARV, 
we assess the consistency $\omega_u$ between the generated pseudo items $\mathcal{S}_u^{pse}$ and the user's original interest patterns. Therefore, $\omega_u$ represents the reliability of the enhanced sequence $\widetilde{\mathcal{S}}_u$.

\subsection{Dual-Channel Training (DCT)}\label{subsec:train}

To integrate the reliability evaluation results from ARV into the training process, we propose a Dual-Channel Training strategy, which consists of two components: weight decay and loss function weighting.

During training, we use binary cross-entropy loss~\cite{kang2018sasrec} as the objective function:
\begin{equation}\label{eq7}
    \mathcal{L}(\mathcal{S}_u)=-\sum_{t=1}^{n_u}\left[log(\sigma(v_t^{+}))
        +log(1-\sigma(v_t^{-}))\right]
\end{equation}
Here, 
$t$ denotes the time step, which corresponds to the 
$t$-th interaction in the user's interaction history, and $n_u$ refers to the number of interactions of user $u$. $v_t^{+}$ represents the positive samples at time step $t$, while $v_t^{-}$ refers to the negative samples at time step $t$. $\mathcal{L}(\mathcal{S}_u)$ represents the value of the BCE loss function computed when the sequence is $\mathcal{S}_u$. 
$\sigma(\cdot)$ represents the sigmoid function.

\noindent\textbf{\textit{Weight Decay}}. 
To adapt data augmentation to SRS training, we weight the enhanced sequence $\widetilde{\mathcal{S}}_u$ according to its reliability $\omega_u$. Given that head users (the users have many interactions) already have an extensive history of interactions for the SRS to learn their interest patterns, we introduce the weight decay module to reduce the weight of the enhancement sequence for head users. This allows LLMSeR to focus more on the user's historical interactions. For tail users, our goal is to ensure that the generated pseudo items effectively alleviate the long-tail user problem. Therefore, the weight of $\widetilde{\mathcal{S}}_u$ for tail users will not be reduced. Formally, the weight decay process is as follows:
\begin{equation}\label{eq9}
    \alpha_u=\left\{ 
    \begin{aligned}
        & \beta\cdot \omega_u & &n_u>T\\
        & \omega_u & &n_u <= T
    \end{aligned}
    \right.
\end{equation}
Here, $\beta$ is a hyperparameter for weight decay. 
$T$ is the threshold used to distinguish long-tail users from head users, while $\alpha_u$ represents the weight assigned to 
$\widetilde{\mathcal{S}}_u$ during training. 

\noindent\textbf{\textit{Loss Function Weighting}}. 
During the training of the sequential recommendation model, both the enhanced interaction sequence and the historical interaction sequence are used as inputs. 
The final loss function for the sequential recommendation model can be expressed as: 
\begin{equation}\label{eq11}
    \mathcal{L}=\sum_{u=1}^{|\mathcal{U}|}\left[(1-\alpha_u)\cdot \mathcal{L}(\mathcal{S}_u)+\alpha_u\cdot \mathcal{L}(\widetilde{\mathcal{S}}_u) \right]
\end{equation}
\section{Experiment}\label{experiments}

In this section, we present the comprehensive experiments conducted to evaluate the performance of LLMSeR, with the aim of addressing the following \textbf{R}esearch \textbf{Q}uestions (\textbf{RQ}):
\begin{itemize}[leftmargin=*]
    \item \textbf{RQ1:} How does LLMSeR compare to existing data augmentation methods? Is LLMSeR universally applicable across different sequential recommendation models?
    \item  \textbf{RQ2:} Is each component of LLMSeR effective?
    \item \textbf{RQ3:} Can LLMSeR help alleviate long-tail user problem? 
    \item \textbf{RQ4:} How does the number of pseudo items generated by the augmentor impact the performance of LLMSeR?
    \item \textbf{RQ5:} How do the hyperparameters of weight decay influence the performance of LLMSeR?
\end{itemize}


\begin{table*}[t]
\tabcolsep=0.08cm 
    \caption{The overall results of competing methods and LLMSeR on three datasets. Bold indicates the best performance, while underlining highlights the highest score among the baselines. ``*'' indicates the statistically significant improvements (\ie two-side t-test with $p<0.05$) over the best baseline.}
    \centering
    \resizebox{\textwidth}{!}{
    \begin{threeparttable}
        \begin{tabular}{cc|cccc|cccc|cccc}
        \toprule[1.5pt]
        \multirow{2}{*}{\textbf{Backbone}} & \multirow{2}{*}{\textbf{Model}} & \multicolumn{4}{c|}{\textbf{Fashion}} & \multicolumn{4}{c|}{\textbf{Book}} & \multicolumn{4}{c}{\textbf{Yelp}} \cr
        \cmidrule{3-14} 
            & & \textbf{H@10} & \textbf{N@10} & \textbf{H@20} & \textbf{N@20} & \textbf{H@10} & \textbf{N@10} & \textbf{H@20} & \textbf{N@20} & \textbf{H@10} & \textbf{N@10} & \textbf{H@20} & \textbf{N@20} \cr
        \midrule
        \multirow{5}{*}{GRU4Rec} & - None & 0.4101 & 0.3226 & 0.4818 & \underline{0.3407} & 0.2956 & 0.1788 & \underline{0.4199} & 0.2102 & 0.6998 & 0.4240 & 0.8339 & 0.4582\cr
        & - ASReP & 0.4086 & 0.2919 & 0.4921 & 0.3128 & \underline{0.3004} & \underline{0.1807} & \underline{0.4199} & \underline{0.2108} & \underline{0.7184} & \underline{0.4395} & \underline{0.8458} & \underline{0.4719}\cr
        & - DiffuASR & \underline{0.4243} & 0.3101 & \underline{0.4935} & 0.3276 & 0.2753 & 0.1624 & 0.3902 & 0.1913 & 0.5931 & 0.3512 & 0.7481 & 0.3904\cr
        & - LLMRec & 0.4100 & \underline{0.3228} & 0.4775 & 0.3397 & 0.2988 & 0.1764 & 0.4181 & 0.2064 & 0.6922 & 0.4134 & 0.8323 & 0.4490\cr
        \cmidrule{2-14}
        & - \textbf{LLMSeR} & \textbf{0.4346}* & \textbf{0.3510}* & \textbf{0.4970}* & \textbf{0.3667}* & \textbf{0.3359}* & \textbf{0.2125}* & \textbf{0.4587}* & \textbf{0.2429}* & \textbf{0.7666}* & \textbf{0.4790}* & \textbf{0.8813}* & \textbf{0.5027}*\cr
        \midrule
        \multirow{5}{*}{Bert4Rec} & - None & 0.3948 & 0.3099 & 0.4740 & \underline{0.3299} & \underline{0.3730} & \underline{0.2351} & \underline{0.4915} & 0.2649 & 0.7110 & \underline{0.4701} & 0.8253 & \underline{0.4991}\cr
        & - ASReP & 0.4145 & 0.2843 & 0.4750 & 0.3038 & 0.3644 & 0.2327 & 0.4812 & 0.2621 & \underline{0.7121} & 0.4404 & \underline{0.8371} & 0.4721\cr
        & - DiffuASR & \underline{0.4159} & 0.2843 & \underline{0.4830} & 0.3032 & 0.3724 & 0.2434 & 0.4851 & \underline{0.2717} & 0.7108 & 0.4635 & 0.8325 & 0.4944\cr
        & - LLMRec & 0.4029 & \underline{0.3107} & 0.4772 & 0.3293 & 0.3322 & 0.1989 & 0.4505 & 0.2286 & 0.7084 & 0.4647 & 0.8225 & 0.4937\cr
        \cmidrule{2-14}
        & - \textbf{LLMSeR} & \textbf{0.4268}* & \textbf{0.3585}* & \textbf{0.4927}* & \textbf{0.3750}* & \textbf{0.4045}* & \textbf{0.2715}* & \textbf{0.5226}* & \textbf{0.3012}* & \textbf{0.7330}* & \textbf{0.4846}* & \textbf{0.8452}* & \textbf{0.5132}*\cr
        \midrule
        \multirow{5}{*}{SASRec} & - None & 0.4066 & \underline{0.3419} & 0.4772 & \underline{0.3595} & \underline{0.3893} & \underline{0.2704} & \underline{0.5062} & \underline{0.2998} & 0.7334 & 0.4769 & 0.8466 & 0.5057\cr
        & - ASReP & 0.4066 & 0.3414 & 0.4723 & 0.3579 & 0.3851 & 0.2610 & 0.4978 & 0.2892 & 0.7425 & 0.4719 & 0.8576 & 0.5011 \cr
        & - DiffuASR & \underline{0.4153} & 0.3355 & \underline{0.4844} & 0.3529 & 0.3789 & 0.2582 & 0.4884 & 0.2857 & \underline{0.7435} & 0.4755 & \underline{0.8598} & 0.5052\cr
        & - LLMRec & 0.4050 & 0.3365 & 0.4788 & 0.3550 & 0.3881 & 0.2650 & 0.5006 & 0.2933 & 0.7390 & \underline{0.4779} & 0.8508 & \underline{0.5064}\cr
        \cmidrule{2-14}
        & - \textbf{LLMSeR} & \textbf{0.4325}* & \textbf{0.3611}* & \textbf{0.4962}* & \textbf{0.3771}* & \textbf{0.4190}* & \textbf{0.2868}* & \textbf{0.5310}* & \textbf{0.3150}* & \textbf{0.7558}* & \textbf{0.4950}* & \textbf{0.8670}* & \textbf{0.5187}*\cr
        \bottomrule[1.5pt]
        \end{tabular}
    \end{threeparttable}
    \label{tab:overall}
    }
\end{table*}

\subsection{Experiments Settings}\label{subsec:settings}

\noindent\textbf{\textit{Datasets}}.
We conduct experiments on three public datasets: Fashion, Book, and Yelp. 
The Amazon\footnote{\url{https://cseweb.ucsd.edu/~jmcauley/datasets.html\#amazon\_reviews}}~\cite{ni2019amazon} dataset is a large e-commerce collection containing user ratings of items and rich item descriptions. 
It includes several sub-datasets, and we primarily use the \textbf{Fashion} and \textbf{Book} sub-datasets for our experiments. The \textbf{Yelp}\footnote{\url{https://www.yelp.com/dataset}} dataset contains user check-in records across various domains, including restaurants, shopping malls, hotels, and more. More details about the datasets can be found in \textbf{Appendix C.1}.


\noindent\textbf{\textit{Baselines}}. \label{subsubsec:baseline}
We compare LLMSeR with co-occurrence-based methods and LLM-based methods in the field of data augmentation. Specifically, \textbf{ASReP}~\cite{ASReP2021augmenting} and \textbf{DiffuASR}~\cite{difuasr2023diffusion} are co-occurrence-based methods, while \textbf{LLMRec}~\cite{wei2024llmrec} is an LLM-based method. We also include a baseline called ``\textbf{None}'' that does not use augmented data to highlight the impact of data augmentation. More details about baselines can be found in \textbf{Appendix C.2}.


\noindent\textbf{\textit{Backbones}}. \label{subsubsec:backbone}
LLMSeR can be seamlessly integrated with most SRS models. To validate its generality, we apply proposed LLMSeR data augmentation method to three backbone models. They are GRU4Rec~\cite{gru4rec2015session}, Bert4Rec~\cite{sun2019bert4rec} and SASRec~\cite{kang2018sasrec}. More details about the backbones can be found in \textbf{Appendix C.3}.

\noindent\textbf{\textit{Implementation Details}}.
All experiments in this paper are conducted on a platform featuring the AMD EPYC 7542 processor and the NVIDIA GeForce RTX 3090 GPUs.
In the experiment, the LLM used by LLMSeR is the GLM-4-Flash model from ChatGLM\footnote{\url{https://chatglm.cn/}}~\cite{zeng2022chatglm}.
The hyperparameter $\beta$ for LLMSeR is searched from the set [$0.1$, $0.2$, $0.3$, $0.4$, $0.5$, $0.6$, $0.7$, $0.8$]. 
More details about the implementation details can be found in \textbf{Appendix C.4}.

\noindent\textbf{\textit{Evaluation Metrics}}.\label{subsubsec:evaluation}
We use two standard metrics to evaluate the performance of LLMSeR in the top-$K$ recommendation task: Hit Rate (H@$K$) and Normalized Discounted Cumulative Gain (N@$K$). The values of $K$ are set to $10$ and $20$.

\subsection{Overall Performance (RQ1)}\label{subsec:overall}


Table \ref{tab:overall} presents a performance comparison of three SRS enhanced by LLMSeR and several competitive baseline methods. The results indicate that the recommendation model using LLMSeR outperforms all baseline models. A more detailed analysis of the experimental results in Table \ref{tab:overall} is provided below to respond the \textbf{RQ1}.

In terms of overall performance, LLMSeR outperforms the baseline methods across all three models, demonstrating the superiority and versatility of our approach. Upon further analysis of the two baseline methods, ASReP and DiffuASR, which are co-occurrence-based, we observe that they do not consistently improve the performance of the backbone models. This may be due to the inaccurate similarity estimation for long-tail users, which causes the pseudo item generation to be influenced by noisy items. As for LLMRec, its performance generally lags behind that of LLMSeR, as it is more susceptible to hallucinations from the LLMs, which may diminish the benefits of data augmentation.

By examining the performance of the three backbones across different datasets, we observe that the improvement brought by LLMSeR to the backbone model varies. The most significant improvement occurs on the Fashion dataset, followed by the Book dataset, and finally the Yelp dataset. This may be related to the average length of user interaction sequences in the dataset. Among the three datasets, the long-tail user problem in fashion data has the largest impact on the SRS, leading to the most significant improvement.

\subsection{Ablation Study (RQ2)} \label{subsec:ablation}

\begin{table}[t]
    \caption{The ablation study conducted on the Fashion dataset and based on the GRU4Rec. The boldface refers to the highest score.}
    \centering
    \begin{threeparttable}
        \begin{tabular}{c|cccc}
        \toprule[1.5pt]
        \multirow{2}{*}{\textbf{Model}} & \multicolumn{2}{c}{\textbf{Overall}} &  \multicolumn{2}{c}{\textbf{Tail}} \cr
        \cline{2-5} 
            & \textbf{H@10} & \textbf{N@10} & \textbf{H@20} & \textbf{N@20} \cr 
        \midrule
        \textbf{LLMSeR} & \textbf{0.4346} & \textbf{0.3510} & \textbf{0.3616} & \textbf{0.2682}\cr
        - \textit{w/o} CCG & 0.4210 & 0.3235 & 0.3424 & 0.2333\cr
        - \textit{w/o} SNF & 0.4222 & 0.3164 & 0.3464 & 0.2325\cr
        - \textit{w/o} ARV & 0.4279 & 0.3422 & 0.3523 & 0.2547\cr
        - \textit{w/o} RCS & 0.4179 & 0.3350 & 0.3392 & 0.2449\cr
        - \textit{w/o} Reason & 0.4189 & 0.3377 & 0.3434 & 0.2500\cr
        - \textit{w/o} WD & 0.4267 & 0.3487 & 0.3546 & 0.2677\cr
        \bottomrule[1.5pt]
        \end{tabular}
    \end{threeparttable}
    \label{tab:ablation}
\end{table}

For \textbf{RQ2}, we conduct ablation experiments using GRU4Rec on the Fashion dataset, and the results are presented in Table \ref{tab:ablation}. 
The ``Tail'' refers to users whose number of interactions $n_u$ does not exceed the threshold $T$, specifically those with no more than $3$ interactions in the Fashion dataset. 
First, we design two variants of the SIA, \ie ``- \textit{w/o} CCG'' and ``- \textit{w/o} SNF''. 
``- \textit{w/o} CCG'' does not incorporate collaborative information in the enhancement process, while ``- \textit{w/o} SNF'' omits semantic information. Both variants show reduced overall performance as well as decreased performance on tail users, highlighting the effectiveness of jointly utilizing collaborative and semantic information for data enhancement.
Then, We analyze the performance of ARV by designing three additional variants. In one variant (- \textit{w/o} ARV), we remove ARV entirely and set the weight of the augmented sequence to $1$ during training. The overall performance of this variant, along with its performance on tail users, is reduced, indicating that the LLM hallucination problem hampers GRU4Rec’s ability to model user preferences effectively. Further, in the ARV module, we observe a decrease in performance when either the LLM Reason (- \textit{w/o} Reason) or the RCS (- \textit{w/o} RCS) component is removed, demonstrating that both components contribute positively to LLMSeR. 
Lastly, we eliminate the weight decay module from the DCT strategy (- \textit{w/o} WD). 
From the results, it can be observed that the performance of this variant on tail users decreases the least compared to the other variants. This suggests that the weight decay module mainly serves to preserve the performance for head users, and its removal primarily affects the performance on head users, leading to a decrease in overall performance. More details about variants can be found in \textbf{Appendix D.1}.

\subsection{Long-tail Problem Study (RQ3)}\label{subsec:group}

\begin{figure}[t]
\centering
\includegraphics[width=1.0\linewidth]{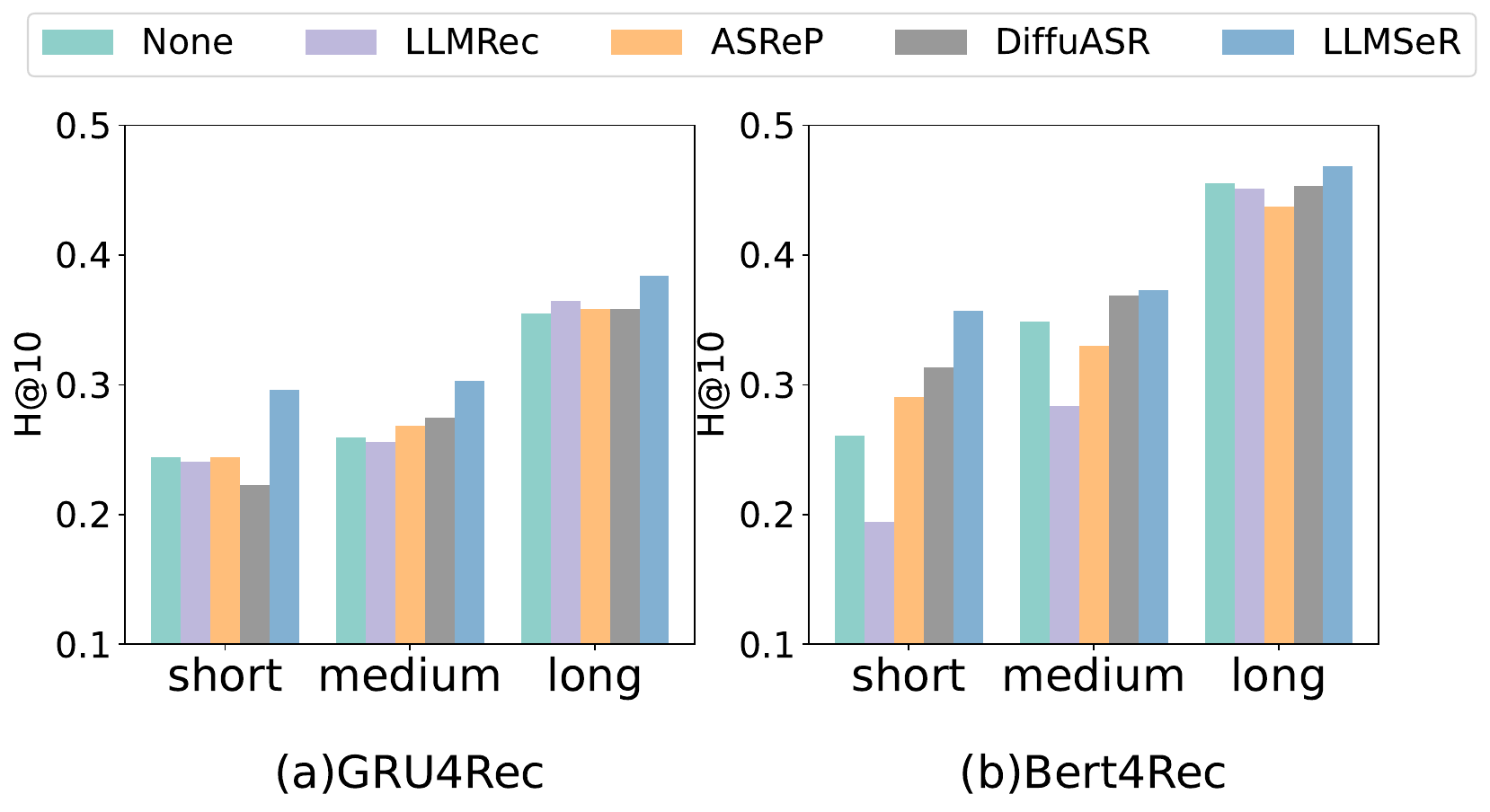}
\caption{Performance on the Book dataset with grouped users. ``short'' refers to users with interaction lengths in the range of $(0, 4)$, ``medium'' refers to users with interaction lengths in the range of $[4, 6)$, ``long'' refers to the range of $[6, \infty)$.}
\label{fig:exp_group}
\end{figure}

\begin{figure*}[!t]
    \begin{minipage}[t]{0.325\linewidth}
        \centering
        \begin{subfigure}{1\linewidth}  \includegraphics[scale=0.55]{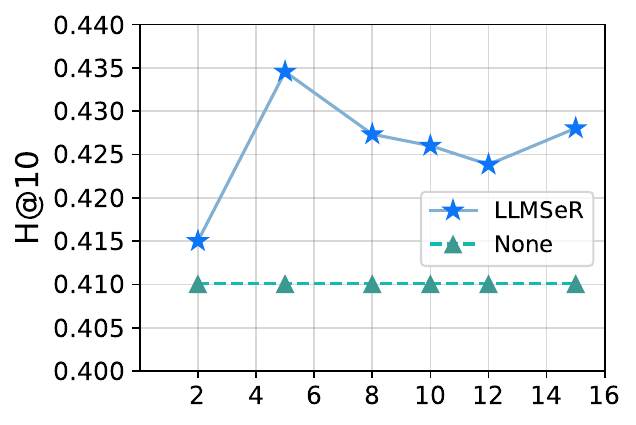}
            \caption{GRU4Rec}
            \label{fig:numPseGRU}
        \end{subfigure}
    \end{minipage}
    \begin{minipage}[t]{0.325\linewidth}
        \centering
        \begin{subfigure}{1\linewidth}
            \includegraphics[scale=0.55]{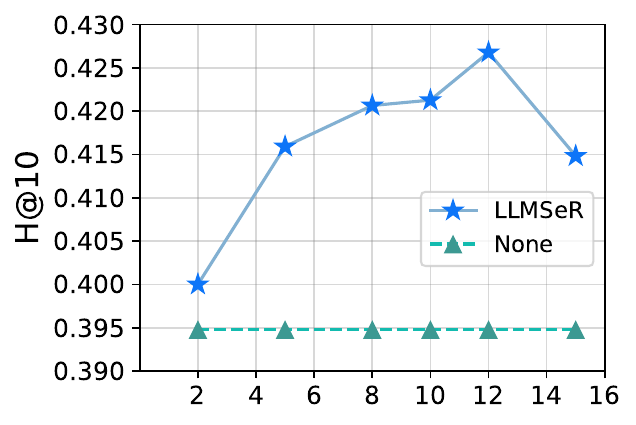}
            \caption{Bert4Rec}
            \label{fig:numPseBert}
        \end{subfigure}
    \end{minipage}
    \begin{minipage}[t]{0.325\linewidth}
        \centering
        \begin{subfigure}{1\linewidth}
            \includegraphics[scale=0.55]{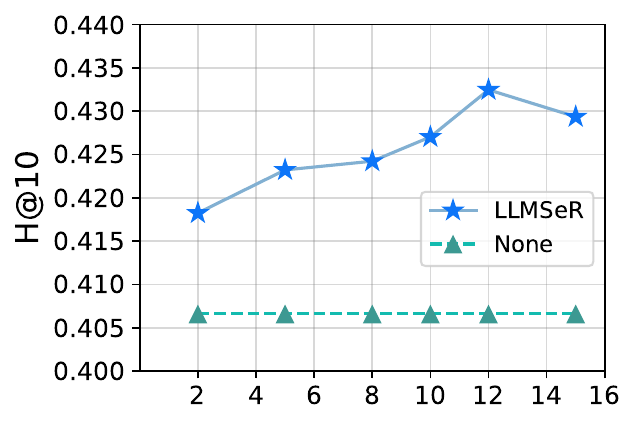}
            \caption{SASRec}
            \label{fig:numPseSAS}
        \end{subfigure}
    \end{minipage}
    \caption{The results of experiments for the number of pseudo items $M$ for each user. All the results are conducted on Fashion dataset.}
    \label{fig:pseNum}
\end{figure*}

To address \textbf{RQ3}, we evaluate the performance of LLMSeR across different user groups, and the results are presented in Figure \ref{fig:exp_group}. As shown, the three backbone models enhanced with LLMSeR outperform the competitive baselines in all user groups, demonstrating the effectiveness and generality of LLMSeR. 
By examining the performance of LLMSeR across the three user groups, we observe that while it enhances the recommendation efficiency for short-group users, the experience of long-group users remains unaffected, and even improves slightly. This indicates that LLMSeR effectively alleviates the long-tail user problem without significantly affecting the experience of head users with rich interaction data. More details about the analysis of different user groups can be found in \textbf{Appendix D.2}.

\subsection{Number of Pseudo Items (RQ4)}\label{subsec:pseudoNum}


To evaluate the impact of the number of pseudo items on LLMSeR mentioned in \textbf{RQ4}, we report the performance of LLMSeR under different numbers of pseudo items on the Fashion dataset for three backbones, as shown in Figure \ref{fig:pseNum}. By observing the performance of Bert4Rec and SASRec, we can see that as the number of pseudo items $M$ increases from $2$ to $12$, the recommendation performance of the backbone models improves. This is because a greater number of pseudo items enriches the user's interaction sequence and helps alleviate the long-tail user problem. However, when the number of pseudo items $M$ exceeds $12$, the recommendation performance begins to decline. This may be because an excessive number of pseudo items is more likely to generate noise, which reduces the reliability of the enhanced sequence $\widetilde{\mathcal{S}}_u$ and leads to a smaller weight for $\widetilde{\mathcal{S}}_u$ during training. In conclusion, it is crucial to select an appropriate number of pseudo items $M$ to maximize the benefits of data augmentation.

\subsection{Hyperparameter Analysis (RQ5)}\label{subsec:parameter}

\begin{figure}[!t]
    \begin{minipage}[t]{0.495\linewidth}
        \centering
        \begin{subfigure}{1\linewidth}
            \includegraphics[scale=0.414]{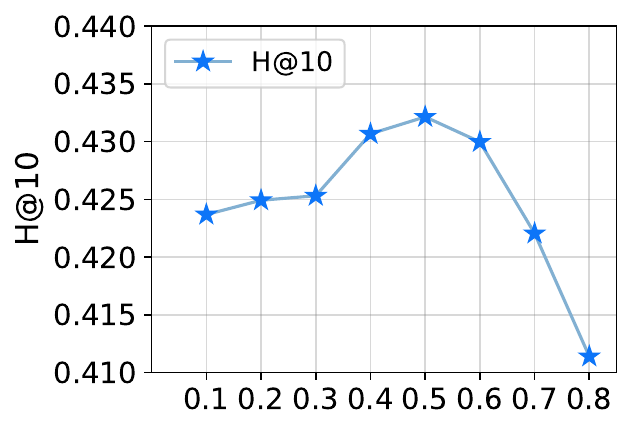}
            \caption{}
        \end{subfigure}
    \end{minipage}
    \begin{minipage}[t]{0.495\linewidth}
        \centering
        \begin{subfigure}{1\linewidth}
            \includegraphics[scale=0.414]{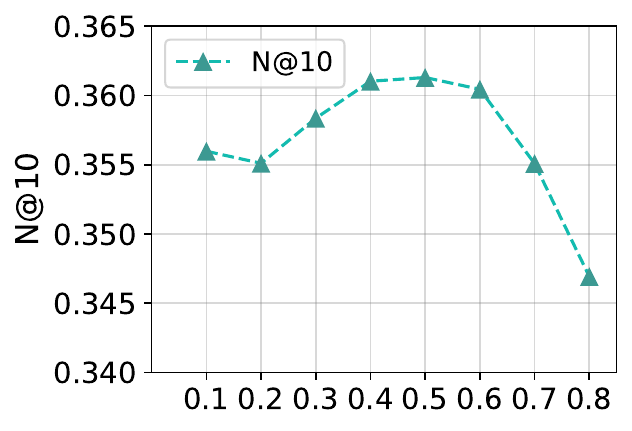}
            \caption{}
        \end{subfigure}
    \end{minipage}
    \caption{Experimental results of weight decay coefficient $\beta$ on the Fashion dataset for SASRec backbone models.}
    \label{fig:beta}
\end{figure}


The weight decay coefficient $\beta$ is a critical hyperparameter in our LLMSeR. To address \textbf{RQ5}, we present the performance trends of SASRec under different values of $\beta$ in Figure \ref{fig:beta}. As $\beta$ increases from $0.1$ to $0.8$, the performance of LLMSeR initially improves and then deteriorates. In the case of SASRec, when $\beta$ increases from $0.1$ to $0.5$, LLMSeR benefits from appropriately learning head users’ preferences through the enhanced sequence $\widetilde{\mathcal{S}}_u$ during training, leading to a more refined modeling of these users. However, when $\beta$ exceeds $0.5$, LLMSeR places excessive emphasis on the enhanced sequence for head users, rather than modeling their already rich historical interactions, which results in a decline in performance. Therefore, selecting an appropriate value for $\beta$ is crucial for the effectiveness of LLMSeR. More details about hyperparameter analysis can be found in \textbf{Appendix D.3}.
\section{Related Works}

\textbf{Sequential Recommendation}. The core goal of sequential recommendation is to predict users' next potential interaction based on their historical interaction sequence~\cite{wang2019sequential}. To extract user preferences from these historical interactions, various deep learning models have been applied, including RNNs~\cite{gru4rec2015session,rnnexample,rnnexample2}, CNNs\cite{tang2018cnn,cnnexample}, and attention mechanisms~\cite{sun2019bert4rec,kang2018sasrec}. 
For example, 
GRU4Rec~\cite{gru4rec2015session} employs GRU layers to model behavioral sequences.
Despite advances in SRS, the long-tail user problem remains a significant challenge that undermines its performance.
Some data augmentation approaches have been proposed to address this challenge~\cite{ASReP2021augmenting,BiCAT2021sequential,difuasr2023diffusion}. Traditional co-occurrence-based methods rely solely on collaborative signals, which are often plagued by noise due to the inaccurate similarities between long-tail users. In this paper, we propose a method that combines semantic and collaborative information to jointly enhance user interaction sequences, aiming to alleviate the long-tail user problem.

\vspace{1mm}
\noindent\textbf{LLMs for Recommendation}. LLMs have demonstrated exceptional capabilities in language understanding~\cite{openai2023gpt} and complex reasoning~\cite{reason2024language}, prompting some research into their application in recommendation systems.~\cite{llmforRS2024survey} These efforts can be broadly categorized into two branches: LLMs as recommendation systems~\cite{wang2023drdt,bao2023tallrec} and LLM-enhanced recommendation systems~\cite{wei2024llmrec,llama4rec2024integrating,huang2024llmins,TedRec2024sequence,llmaug2024breaking}. 
In the first category, DRDT~\cite{wang2023drdt} enables LLMs to explore, critique, reflect, and ultimately model user preferences through prompts. 
However, these approaches may encounter challenges, such as an increase in inference time for SRS~\cite{liu2024LLMAug}. Another branch of methods enhances RS using LLMs without involving them in the inference stage, effectively avoiding this issue. As a result, LLM-based methods for data enhancement in SRS have emerged. Nevertheless, these methods~\cite{wei2024llmrec,llama4rec2024integrating} often lack collaborative signals and are vulnerable to the hallucination problem inherent in LLMs. To address these challenges, we propose LLMSeR, a method that incorporates collaborative signals into LLM-based frameworks and alleviates the impact of hallucinations, thereby improving the overall effectiveness of SRS.
\section{Conclusion}

In this paper, we propose a large language models empowered data augmentation framework
for Sequential Recommendation (LLMSeR). To handle the issue of missing collaborative signals, we design the Semantic Interaction Augmentor. Additionally, to alleviate the impact of LLM hallucination on SRS, we propose the Adaptive Reliability Validation. Furthermore, we design a Dual-Channel Training approach to adapt data augmentation for SRS training. LLMSeR can be seamlessly integrated with most SRS models. The experimental results on three popular SRS models highlight the generality and effectiveness of LLMSeR.


\bibliographystyle{named}
\bibliography{cite}

\end{sloppypar}
\end{document}